\documentclass{llncs}

\usepackage{graphicx}
\usepackage{enumitem}
\usepackage{amsmath}
\usepackage{subfig}
\usepackage{breqn}
\usepackage{pbox}
\usepackage{appendix}
\usepackage{multirow}
\usepackage{url}
\usepackage{flushend}
\usepackage{caption}

\usepackage{setspace}
\usepackage[linesnumbered,ruled,vlined]{algorithm2e}
\SetAlFnt{\scriptsize}
\usepackage{listings}

\usepackage{color}
\definecolor{bluebell}{rgb}{0.64, 0.64, 0.82}
\definecolor{gray}{rgb}{0.4,0.4,0.4}
\definecolor{darkblue}{rgb}{0.0,0.0,0.6}
\definecolor{cyan}{rgb}{0.0,0.6,0.6}

\usepackage{listings,xcolor}
\lstdefinelanguage{XML}
{
  morestring=[b]",
  morestring=[s]{>}{<},
  morecomment=[s]{<?}{?>},
  stringstyle=\color{black},
  identifierstyle=\color{darkblue},
  keywordstyle=\color{cyan},
  morekeywords={vnfCertificate}% list your attributes here
}

\begin{document}
\title{Towards a Trust Aware Network Slice based End to End Services for Virtualised Infrastructures}

\author{%
Vijay Varadharajan\inst{1} \and Kallol Karmakar \inst{1}\and Uday Tupakula\inst{1} \and Michael Hitchens\inst{2} }%
\institute{
Advanced Cyber Security Research Centre, The University of Newcastle, Australia\\\
\email{[vijay.varadharajan, kallolkrishna.karmakar, uday.tupakula]@newcastle.edu.au}
\and
Macquarie University, Australia\\
\email{michael.hitchens@mq.edu.au} }
\maketitle
\begin{abstract}
Future communication networks such as 5G are expected to support end-to-end delivery of services for several vertical markets with diverging requirements. Network slicing is a key construct that is used to provide end to end logical virtual networks running on a common virtualised infrastructure, which are mutually isolated. Having different network slices operating over the same 5G infrastructure creates several challenges in security and trust. This paper addresses the fundamental issue of trust of a network slice. It presents a trust model and property-based trust attestation mechanisms which can be used to evaluate the trust of the virtual network functions that compose the network slice. The proposed model helps to determine the trust of the virtual network functions as well as the properties that should be satisfied by the virtual platforms (both at boot and run time) on which these network functions are deployed for them to be trusted. We present a logic-based language that defines simple rules for the specification of properties and the conditions under which these properties are evaluated to be satisfied for trusted virtualised platforms. The proposed trust model and mechanisms enable the service providers to determine the trustworthiness of the network services as well as the users to develop trustworthy applications. We have developed a trust management architecture and mechanisms that enable the service providers to determine the trustworthiness of the network slices providing the network services.  We have implemented a prototype of the trust management architecture using Open Source MANO Platform and presented the performance results. The results show that our trust mechanisms cause only a slight delay in the performance of the deployment of network slices establishing end to end services over virtualised infrastructures. We have also discussed how the proposed architecture can be used to detect and mitigate the impact of malicious virtual network functions in a dynamic manner.
\end{abstract}
\begin{keywords}
Network Function Virtualisation Trust, Network Slice Trust, Trust Management, Property-based Attestation.
\end{keywords}
\vspace{-0.2cm}
\section{Introduction}
\vspace{-0.3cm}
\noindent Future communication networks such as 5G are expected to support end-to-end delivery of services for several vertical markets with diverging requirements. 5G networks will provide virtually ubiquitous, ultra-high bandwidth and low latency connectivity not only to individual users but also to connected objects. Unlike the previous network architectures, 5G architectures are programmable to accommodate user requests at run-time, allowing network and providers to derive much benefit through high network throughput and performance~\cite{navarro2020survey}.\\
Software Defined Networks (SDN) and Network Function Virtualisation (NFV) are the two key technologies that fuel the development of 5G architecture~\cite{abdelwahab2016network}. SDN enables programming and control of virtualized network resources, whereas NFV allows network functions to be virtualized and run over commodity hardware, enabling flexible resource allocation and sharing. The combination of SDN and NFV helps to make 5G an elastic, resource agnostic, service-oriented network. European Telecommunications Standards Institute (ETSI) has developed a 5G network architecture standard known as open source Management and Orchestration (MANO)~\cite{etsi2013network}. It introduces a key concept called network slicing to create end to end logical virtual networks running over a common virtualised infrastructure, which are mutually isolated. The SDN and NFV technologies provide the programmability and flexibility to create these network slices, enabling service providers and end users (customers) to request, deploy and provision end-to-end network services on-demand without concerning about the underlying hardware. However having different network slices operating over the same 5G network infrastructure creates several challenges in security ~\cite{lal2017nfv},~\cite{yang2016survey},~\cite{benzaid2020zsm},~\cite{bursell2014network},~\cite{etsi2017v3},~\cite{etsi2016003}. 
%However, it has created new attack surfaces due to the use of hypervisors for the virtual network functions~\cite{lal2017nfv}. 

\noindent Trust is a fundamental challenge in the design and deployment of network slices, as they can belong to different users and service providers, and can be implemented over different infrastructures provided by different operators, providing services to applications in different business sectors~\cite{etsi2017v3},~\cite{etsi2016003}. A key design issue is whether a particular platform in the infrastructure (e.g. a virtual machine) is in a state that it is expected to be, that is, a trusted state before the deployment of virtual network functions. Moreover, as networks and virtual machine platforms such as hypervisors are susceptible to be a range of security attacks, these infrastructure components can become untrusted during the course of operation, impacting the network services that are being run on them. Furthermore, a network slice may use some virtual network functions developed by third parties, which can be malicious or can get infected during run time. Therefore the ability to reason about trust of network slices is critical, both from the user and provider perspectives, when it comes to reasoning about trust of 5G networks and the provisioning of secure network services. Hence the focus of this paper is to develop techniques for reasoning about trust properties of network slices which form the basis of provision of secure end to end services over 5G networks.\\
\noindent We develop a trust model and the trust properties that need to be satisfied by the various components in the infrastructure in the creation to deployment of trusted network slices leading to establishment of 5G network services. Our trust model specifies trust relations between the different 5G components and actors in terms of property based attestation properties of virtualised platforms providing the network services. We introduce a logic based language to capture the trust requirements and specify the trust relations and properties in the 5G network infrastructure. Then we develop a trust management architecture to evaluate the trust of the virtual network functions being deployed in virtual machines and determine the trustworthiness of network slices providing end to end services in 5G networks.\\
\noindent The significance of this work is that the proposed trust model and architecture can be used to enhance the quality of decisions made by the services in the 5G architecture, including security services such as access control and attack detection decisions. This in turn can lay the foundations of a trust enhanced security architecture for 5G network services.\\
\noindent The paper is organized as follows. First, in Section 2, we begin with a brief description of the architecture of the 5G virtualised network infrastructure. This will help to identify explicitly the basic trust issues that can arise in such an environment. Section 3 specifies the trust model and the trust properties that need to be satisfied by the various components in the architecture that are involved in the creation and deployment of network slices. We also introduce a logic based policy language that is used  to specify the trust properties and the trust derivation process to evaluate the trust of a network slice using the trust model. Section 4 describes our prototype implementation of the trust model and instantiation of trusted network slices using the Open Source MANO platform for 5G networks. In Section 5, we present an analysis of trust properties of our prototype as well as an analysis of network performance and their results. In Section 6, we compare our trust model and architecture with previous relevant related works. Finally, Section 7 summarises our conclusions and outlines some future work.

\vspace{-0.2cm}
\section{Architecture Overview and Network Slice Deployment}
\vspace{-0.3cm}
In this section, we briefly outline the key components of the 5G network architecture that is directly relevant to the trusted network slice discussions in this paper. The purpose of this architectural description is that it will enable us to extract the trust requirements and assumptions needed for the design of the trust model and mechanisms in the next section.

\noindent \underline{\bf{Architecture Overview}}

\vspace{0.1cm}

\noindent In this network architecture~\cite{nfv2017001}, software-based virtual network functions (VNFs) are instantiated across a diverse range of virtual machines (VMs), connected and chained together in a certain way to achieve the desired network services. This architecture is mainly composed of three main functional blocks: network function virtualisation infrastructure (NFVI), VNFs, and NFV Management and Orchestration (NFV MANO), as illustrated in Fig. 2.  Associated with these blocks are their respective management authorities, namely Virtualized Infrastructure Manager (VIM), VNF Manager (VNFM) and  NFV Orchestration Manager (NFVO-M). 

\noindent Consider a simplified view of such an architecture. The VNFs are implemented as software instances running on infrastructure resources, which include hardware, storage capacity, networking resources and physical assets for radio access. These VNFs are managed by a VNFM. The virtual and physical resources on which these VNFs run are managed by the VIM in the domain. NFVO-M manages both resource and service orchestration. The resource orchestration involves the coordination of resources under different VIMs, whereas service orchestration involves the creation and coordination of end-to-end services with VNFs under different VNFMs.

\noindent \underline{\bf{Network Slice Deployment}}

\vspace{0.1cm}
\noindent The virtualised network functions and the associated authorities enable the 5G infrastructure to support network slicing. A network slice comprises a collection of VNFs to form the network services that are delivered to the users (or tenants). Each tenant can have several network slices under its management and a network slice can be used by several tenants in a multi-tenant environment. Network slicing enables concurrent logical networks providing various network services and functionalities to different tenants. For instance, an emergency network slice that allows immediate and on-demand services for patients, a highly secure communication slice for confidential video and audio communications for defense users, and a low-bandwidth network slice for IoT devices used by IoT users and applications.

\noindent The creation and deployment of network slices use their associated resource and infrastructure attributes; NFVO-M maintains a repository of these network slices with their related attributes.
The VNFs in a network slice are deployed in virtual machines (VMs) running over infrastructure provided by the cloud or network provider. VNFM deploys the VNFs on VMs that are suitable, depending on VMs' attributes. The VM resources are managed by the VIM through SDN controller. The SDN controller works in conjunction with the VIM in the creation and establishment of both physical and virtual network communications. Finally, the users are assigned to these created and deployed network slices providing them with the requested network services.
\vspace{-0.4cm}
\subsection{Problem Motivation and Trust Requirements}
\vspace{-0.2cm}
Network service providers or users (tenants) request the creation and deployment of network slices. As outlined above, authorities take the necessary steps to create, deploy and provision the network slices to the requester. The requester uses the network services provided by the network slices for their information transfers and their applications. This raises the fundamental issue of trust in the provision of network services. That is, the trustworthiness of the network slices and the trustworthiness of the deployment of virtual network functions (VNFs) in the network slices on virtual and physical platforms (VMs) in the infrastructure providing these services. From the tenant or user perspective, how to trust the network slice and the VNFs, NFV infrastructure (NFVI) and their providers? From the point of view of VNF manager, how to trust the platform where VNFs are being deployed and executed? For the infrastructure provider VIM (and SDN Controller), how to trust the VNFs that are being deployed on their resources?
\begin{figure}[!ht]
    \centering
    \includegraphics[scale=.45]{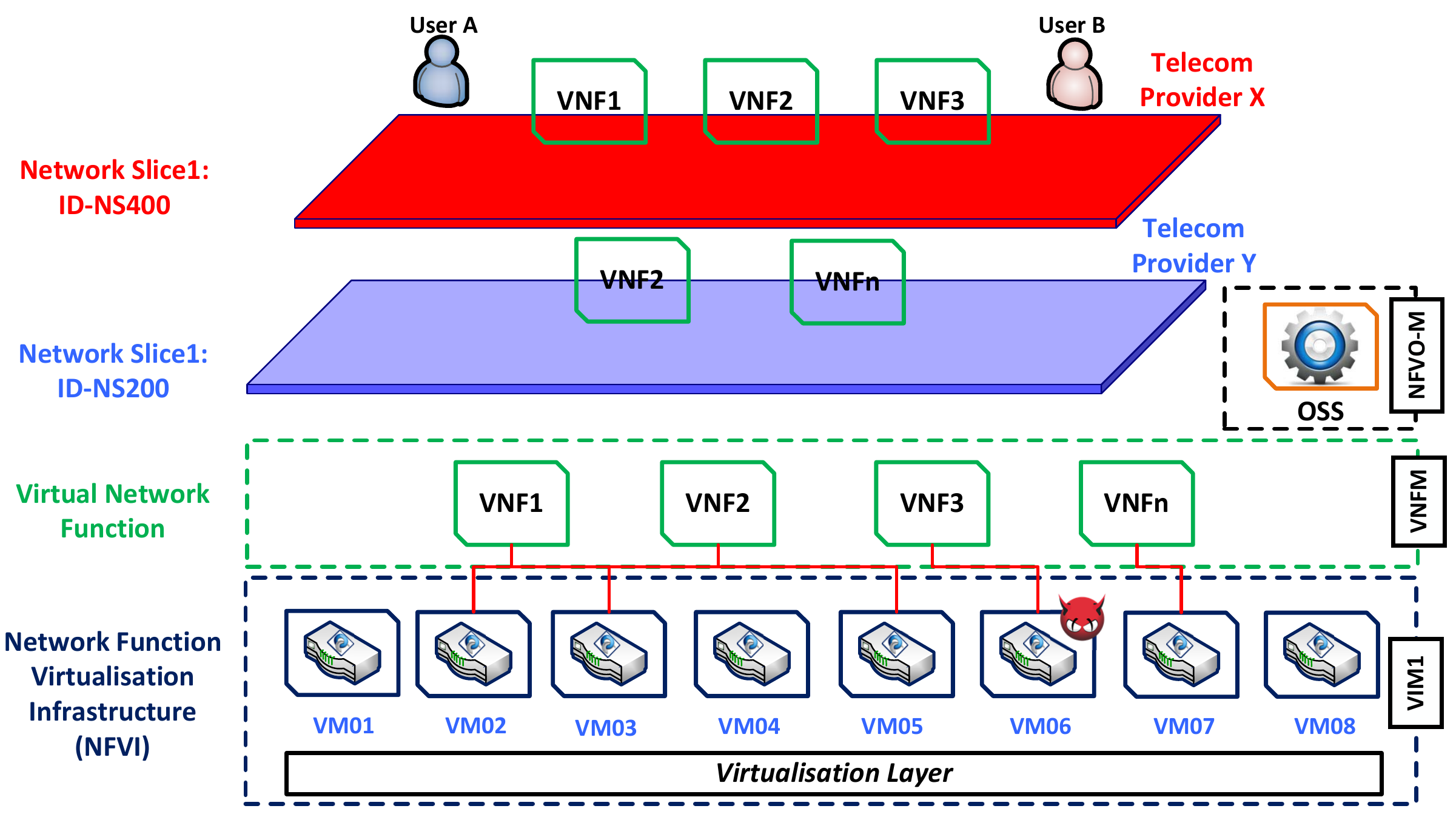}
    \caption{A Network Slice Scenario }
    \label{fig:nssce}
\end{figure}
\noindent There can be a number of security threats on these different infrastructure components such as VNFs, the VMs, the hypervisor on which these VMs run as well as the networks. Furthermore, there can be several parties providing the VNFs and the VMs in the network and cloud infrastructure. Compromise of VMs by an attacker in turn affects the virtual network functions being executed in these VMs. For instance, a logic-bomb integrated into a VM can remain undetected during VNF deployment until activated by a trigger, making the network slice malicious. This will impact the trust of the service provider or the tenant using the services provided by the network slice.\\
Fig~\ref{fig:nssce} shows a scenario in which a malicious virtual machine-VM06 has been used to create a VNF3.  The VM has been tampered during the build process and injected with a logic-bomb. The VNF3 serves as a fast software switch and deployed by the NS400 network slice, which provides a high bandwidth network to users A and B, and the Telco provider X. VNF3 passes all the security checks as the malware code is not active initially. Once deployed the malware in VNF3 only becomes active after it receives a trigger from a specific IP address. Then the software switch starts to perform various malicious activities, making the whole network slice untrustworthy. However, the users A and B, and the Telco provider X are unaware of the malicious behaviour of the network slice. This example illustrates the need for mechanisms that can help the user of the network slice to determine the trustworthiness of a network slice at any time (both at boot time as well as at run time).\\ 
More specifically, the architecture should provide mechanisms that can determine the trust of the various infrastructure components that are used in the creation of the network slice as well as in its deployment on virtual and physical resources in the network infrastructure. That is, in order to reason about the trustworthiness of a network slice, the architecture should provide the following mechanisms: (a) Mechanisms to determine the trust of a VNF at the time of formation of a network slice, (b) Mechanisms to determine the trust of a VM at the time of deployment of VNF on the VM, and (c) Mechanisms to determine the trust of a physical platform before the deployment of the VM on the platform.
\vspace{-0.3cm}
\section{Trust Model}
\vspace{-0.3cm}
In this section, we propose a trust model, which defines the trust properties that need to be satisfied by the the various components such as VNFs,and VMs that are involved in the creation and deployment of network slices. This will then lead to the formulation of the trust of a network slice. The trust management architecture, corresponding to the trust model providing mechanisms identified above (a) to (c), is described in Section 4.\\ 
\noindent Our trust model is based on both binary attestation and property based attestation. In attestation, the basic idea is that an entity is trusted if it behaves as expected for a specific purpose~\cite{mitchell2005trusted}. Trust is evaluated by measuring the state (and behaviour) of an entity at a given point in time, and then comparing the measured state with the (reference) expected state. If the measured state matches with the expected state, then the entity is trusted. \\
In our case, a network slice consists of a set of VNFs that provide the network services of that slice. So a network slice is trusted at a given time if all the VNFs of the slice are trusted at that time. A VNF is trusted if the state of the VNF at that time matches with the reference state of that VNF. As each VNF is a software component, the state of a VNF can be defined in terms of its hashed digest. So by comparing the hashed digest of a VNF (at a given time) with the reference (expected) hashed digest of that VNF provided by the manufacturer, the trust of a VNF can be determined. Hence we can determine the trust of the network slice composed of a set of VNFs, using such hash based binary attestation.  \\
The algorithms required for this attestation include an algorithm to calculate the hashed digest of a VNF and an algorithm to compare this measured value with the reference value. Such matching condition is defined in terms of trust constraints that should be satisfied by a VNF before it is included in the composition of a network slice. Furthermore, these hash values need to be signed by appropriate trusted authorities for them to be trusted (e.g. a Certification Authority). We will consider the trusted authorities in Section 4 on the trust management architecture. \\
However binary hash measurements of a VNF is mainly a representation of the VNF's implementation rather than about its behavioural properties, which are often more important when evaluating trust.  Unlike hash measurements which change whenever a VNF is updated, its properties may not. As properties do not reveal the implementation details of the VNF, they can provide a higher level of privacy. Furthermore, properties can be more meaningful and easier to understand. Hence our model uses property based attestation to reason about the properties satisfied by VNFs and VMs at a given state.\\
In our model, the properties are represented at different levels of granularity. Some properties represent implementation constraints that are enforced on the VNFs such as "absence of buffer overflow". Some others are generic such as a "VNF does not have any third-party software component running within it". In general, our approach is to specify properties as constraints or rules in a policy that need to be satisfied by an entity for it to be trusted. For example, a VNF satisfies the mandatory access control property by enforcing the Bell-La Padula security rules~\cite{bell1973secure}. Or a VNF satisfies the confidentiality property with its output data encrypted using AES algorithm. As in the case of binary hash attestation, we require trusted authorities in property based attestation who can vouch for the properties possessed by an entity. In general, there can be multiple such authorities vouching for different properties. For the specifications of trust properties, we have developed a policy language introduced in the next section (Section 4). 
\vspace{-0.3cm}
\subsection{High Level Trust Model Description}
\vspace{-0.2cm}
We give a brief overview of the trust model at a high level, outlining the preliminaries and basic sets and predicates to that formulate properties associated with components (such as VNFs and VMs) as well as the composition of these components in the form of a network slice. We also outline the trust derivation algorithm to evaluate the trust of the network slice. We have introduced constructs of a simple Logic based language for Property Attestation based Trust (LOPAT) to specify the properties in our trust model. Though these constructs have been in terms of network slices and software components (VNFs), they are equally applicable to systems and hardware components as well as platforms such as hypervisors and virtual machines as the components running on their top.
\begin{definition}
\textbf{Network Slice} - A network slice is a collection of all components (e.g. VNFs) within it.
\end{definition} 
\begin{definition}
\textbf{Component} - Each VNF component inside a Network Slice is termed a component. Each component can perform one or more network functions, either individually or when combined with one or more components in the network slice.  
\end{definition}
\noindent Note such a definition also applies to a hardware component in the system model.
\begin{definition}
\textbf{Composition} - The term composition refers to the act of interconnecting two or more components in a network slice in a certain fashion to provide a specific network service.
\end{definition} 
\noindent In general, the process of composition is an iterative one until the entire network slice is built. It is easier to verify individual components' properties (which are smaller in size) rather than attempting to verify one large monolithic system. Furthermore, the composition process may not preserve the properties of the individual constituent components. This is not a trivial task. 
\begin{definition}
\textbf{Property} - A property is defined as any attribute, characteristic or behaviour associated with a component or a network slice. 
\end{definition} 
\begin{definition}
\textbf{Target:} A target is a subject (e.g. user) or an object to which a property of a component or a network slice is bound. 
\end{definition} 
\noindent For example, let us consider an authentication service where all users are authenticated using passwords and all login attempts are recorded. Here, the authentication component supports the identity verification property of all users (targets). Additionally, the authentication component securely logs (property) all login outcomes (target). Here, we can see that a property of a component is more meaningful when it is bound to a specific target.\\

\noindent The basic preliminaries in our model are as follows:
\begin{enumerate}
\vspace{-0.1cm}
\item 
\textbf{Constant Symbols:} Every member of $NS ~ \cup ~ C  ~ \cup ~ P  ~ \cup ~ T  ~ \cup ~ R  ~ \cup ~ A ~ \cup ~ PER ~ \cup $ $ N $ is a constant. Here, $NS$ is a set of network slices, $C$ is a set of components, $P$ is a set of properties, $T$ is a set of targets, $R$ is a set of resources, $A$ is a set of actions, $PER$ is a set of permissions and $N$ is a set of natural numbers. Each member of the set is represented in the form $set_i$ where $i$ is a member of the set $NS$. For example, $ns_1$, $ns_2$ ........$ns_n$ represent the members of the set $NS$. 
\item \textbf{Variable Symbols:} We define eight sets of variable symbols $V_{ns}, ~V_c, ~V_p, ~V_t,\\ V_o, ~V_a, ~V_{per}$ ranging over the sets $NS, C, P, T, R, A, PER$ respectively. 
\item \textbf{Predicate Symbols:} The following predicate symbols are defined in our model. Each n-ary predicate symbol is represented as p($t_1$,$t_2$,...$t_n$) where p is the name of the predicate and n is its arity.
\begin{enumerate}
\item \textbf{HasC} is a binary predicate symbol. Both the arguments of  HasC are members of $C$. The predicate defines a relationship between two components in a given platform. For example, $HasC(c_1,c_2)$ is read as component `$c_1$' has (or contains) the component `$c_2$'. 
\item \textbf{HasNS} is a binary predicate symbol. The first argument is a member of $NS$ and the second argument is a member of set $C$. It defines a relationship between a network slice and a component. For example, $ {HasNS(ns_1,c_1)}$ is be read as a network slice $ns_1$ has the component $c_1$.
\item \textbf{SatC} is a binary predicate symbol. The first argument is a member of $C$ and the second argument is a member of $P$.  It defines the relationship between a component and the property it satisfies. For example, $SatC(c_1, p_1)$ is read as component $c_1$ satisfies the property $p_1$.
\item \textbf{SatNS} is a binary predicate symbol. The first argument is a member of $NS$ and the second argument is a member of $P$. It defines the relationship between a network slice and the property it satisfies. For example, $SatNS(ns_1, p_1)$ is read as network slices $ns_1$ satisfies the property $p_1$. 
\item \textbf{PreReq} is a quaternary predicate symbol. PreReq defines the `prerequisite' relationship between two SatC predicates. For example,\\ $PreReq((c_1,p_1),(c_2,p_2))$ is equivalent to $\neg SatC(c_1, p_1) \leftarrow \neg SatC(c_2, p_2)$.
\item \textbf{Do} is a quaternary predicate symbol. The first argument is a member of NS, the second argument is a member of R, the third argument is a member of A and the fourth argument is a member of the set Permissions, Per = \{allow,deny\}. It defines the authorisations that are held for each platform on each object. For example, $Do(pf_1,r_1,a_1, allow)$ is read as platform $pf_1$ is allowed to perform action $a_1$ on the resource $r_1$. 	
\end{enumerate}
\end{enumerate}
\begin{definition} \label{def:CP}
\textbf{Component-Property (CP) Rule:} A CP rule is of the form:
\end{definition}
\begin{flushleft} $SatC(c,p) \leftarrow L_1 \wedge L_2 \wedge....L_n$.  \end{flushleft}
\noindent where c,p are elements of the sets C and P respectively, n $\geq$ 0, $L_1$....$L_n$ are either SatC, HasC literals. (i) A rule with only SatC literals in the body indicates the abstraction of one property type to another. There must be at least one SatC literal in the body for this rule. Examples (a) and (b) illustrate this. (ii) The rule may also be used to show that a component may satisfy one property if one of its sub-components satisfies the same (or a different) property. Here, one HasC literal and a minimum of one SatC literal is required. This is shown in example (c) below. 
\begin{flushleft}(a) $ SatC(c_1,p_1) \leftarrow SatC(c_1,p_2)$

(b) $SatC(c_1,Trusted\_True) \leftarrow SatC(c_1,Hash\_e2c182bbb85c2e3a5fcae1936c5900cf91dd7743) \wedge SatC(c_1,Malware\_False)$

(c) $ SatC(c_1,p_1) \leftarrow SatC(c_2,p_2) \wedge HasC(c_1,c_2) $ \end{flushleft}
\noindent The first rule reads: for a given component $c_1$ to satisfy a property $p_1$, $c_1$ must satisfy the property $p_2$.  This rule is particularly useful for abstracting properties at one level to properties at a different level. \\
The second rule reads: component $c_1$ may be trusted (property name = trusted and property value = true) if $c_1$ measures up to the given 160-bit binary value and if $c_1$ has no malware (property name = malware, property value = false). The third rule reads: for a component $c_1$ to satisfy a property $p_1$, $c_2$ must satisfy $p_2$ and $c_1$ must contain $c_2$. For example, an application ($c_1$) can perform auditing operations ($p_1$) if it ($c_1$) contains another sub-component ($c_2$) that records all audit logs ($p_2$). 
\begin{definition} \label{def:NSP}
\textbf{Network Slice-Property (NSP) Rule:} A NSP rule is:
 \end{definition} 
\begin{flushleft} $SatNS(ns,p) \leftarrow L_1 \wedge L_2 \wedge....L_n$. \end{flushleft}
\noindent where ns, p are elements of the sets NS and P respectively, n $\geq$ 0, $L_1$....$L_n$ are either SatC, SatNS or HasNS literals. (i) A rule with only SatNS literals in the body indicates the abstraction of one property type to another. There must be at least one SatNS literal in the body for this rule. Example (a) illustrates this. (ii) The rule may also be used to show that a platform may satisfy one property if one (or more) of its components satisfies the same (or a different) property. Here, one HasNS literal and a minimum of one corresponding SatC literal is required. This is shown in example (b) below. Note that this rule may also contain SatNS literals as shown in example (c).

\begin{flushleft}
	
(a) $ SatNS (ns_1,p_1) \leftarrow SatNS (ns_1,p_2)$

(b) $ SatNS(ns_1,p_1) \leftarrow SatC(c_2,p_2) \wedge HasNS(ns_1,c_2) $ 

(c) $ SatNS(ns_1,p_1) \leftarrow (SatC(c_2,p_2) \wedge HasNS(ns_1,c_2)) \wedge (SatNS(ns_1,p_3)$
 
\end{flushleft}

\noindent Rule (a) reads: A network slice $ns_1$ satisfies the property $p_1$ if the network slice also satisfies property $p_2$. For example, a network slice ($ns_1$) is trustworthy ($p_1$) if it belongs to subnet x ($p_2$). \\
Rule (b) reads: a network slice $ns_1$ is said to satisfy a property $p_1$ if it has a component $c_2$ and if $c_2$ satisfies the property $p_2$. For example, a network slice is considered to be safe ($p_1$) if it has an antivirus software ($c_2$) that is up-to-date ($p_2$).\\
Rule (c) adds an extra condition to Rule (b) that the network slice must also satisfy the property $p_3$. For example, a network slice $ns_1$ is considered to be safe ($p_1$) if it has an antivirus software ($c_2$) that is up-to-date ($p_2$) and if the network slice is spyware-free ($p_3$). 
\vspace{-0.4cm}
\subsection{Trust Evaluation}
\vspace{-0.1cm}
Our trust evaluation is a logic program, which comprises a set P of program clauses. The set P consists of a set of rules and set of facts. The program together with a query Q describes the deduction. Facts are predicates as defined in the previous section. Rules are either CP or NSPP  as given in Definitions~\ref{def:CP} and \ref{def:NSP}. A query is a request for trust derivation accompanied by network slice credentials. The trust resolution ruke determines whether the query is satisfiable (and hence the goal achievable) given the set of facts and rules.

\noindent Now let us consider the trust derivation process from the network slice credentials provided at the time of query request. From the previous section, we use hashed digest certificates and property certificates as credentials.
\begin{enumerate}
\item \textbf{HasNS}: The trust  resolution derives the HasNS information from the hashed digest report generated at the time of attestation request. It includes the list of individual components that are measured in a network slice and their respective hash measurements. The trust resolution uses this report to determine the list of components available in a network slice and builds a list of derived facts in the form of HasNS literals. These derived facts will then be used in the resolution algorithm.
\item \textbf{HasC}: The trust resolution derives the HasC information using the hashed digest report. A HasC predicate defines the decomposition of a component into many sub-components. The hashed digest report contains a list of sub-components for every component measured. The trust resolution can determine which sub-components are included in a given component and builds a list of derived facts in the form of HasC. These derived facts will be used in the trust derivation algorithm.
\item {\textbf{SatPF and SatC}}: The trust resolution derives the properties satisfied by a network slice and component from the property certificates and maps this information to SatNS and SatC literals respectively. These derived facts will be used in the trust derivation algorithm.   
\end{enumerate}
\vspace{-0.4cm}
\subsection{Trust Derivation Algorithm}
\vspace{-0.1cm}
The trust derivation algorithm checks for each SatNS in the query request, whether it can be derived from the network slice property certificate. It also verifies that any prerequisites for SatNS (as facts) are satisfied before marking the SatNS as satisfied. If the SatNS cannot be derived from the property certificates, it checks whether there are NSP rules with SatNS in the head of the rule. It tries to resolve the NSP rule to determine whether SatNS can be satisfied. \\
Then for each HasNS in the NSP rule, it checks whether HasNS can be derived from the hashed digest report. If yes, it marks HasNS as satisfied. Then for each SatC in the NSP rule, it checks whether the SatC can be derived from the property certificates. If yes, it marks SatC as satisfied. If not, it carries out CP resolution.\\
The evaluation for CP resolution is similar to the above. For each SatC in the CP rule, it first checks whether the SatC can be derived from the component property certificate. If so, it also verifies that any prerequisites for SatC (as facts) are satisfied before marking SatC as satisfied. If the SatC cannot be derived from the property certificates, it checks whether there are CP rules with SatC as head of the rule. It tries to resolve the CP rule to determine whether SatC can be satisfied.\\
Then for each HasC in the CP rule, it checks whether HasC can be derived from the property certificates. If yes, it marks the HasC literal as satisfied. If all literals are satisfied, then it goes back to the return to query request resolution.   	
Then finally, if all literals required in the query request are satisfied, then the algorithm outputs the trust outcome. 
\section{Trust Management Architecture (T-MANO)}
\vspace{-0.3cm}
In this section, we describe the trust management architecture and the prototype implementation instantiating trusted network slices using the Open Source MANO platform for 5G networks.
\vspace{-0.3cm}
\subsection{Trust Management Architecture}
\vspace{-0.2cm}
\begin{figure}[!ht]
    \centering
    \includegraphics[scale=.54]{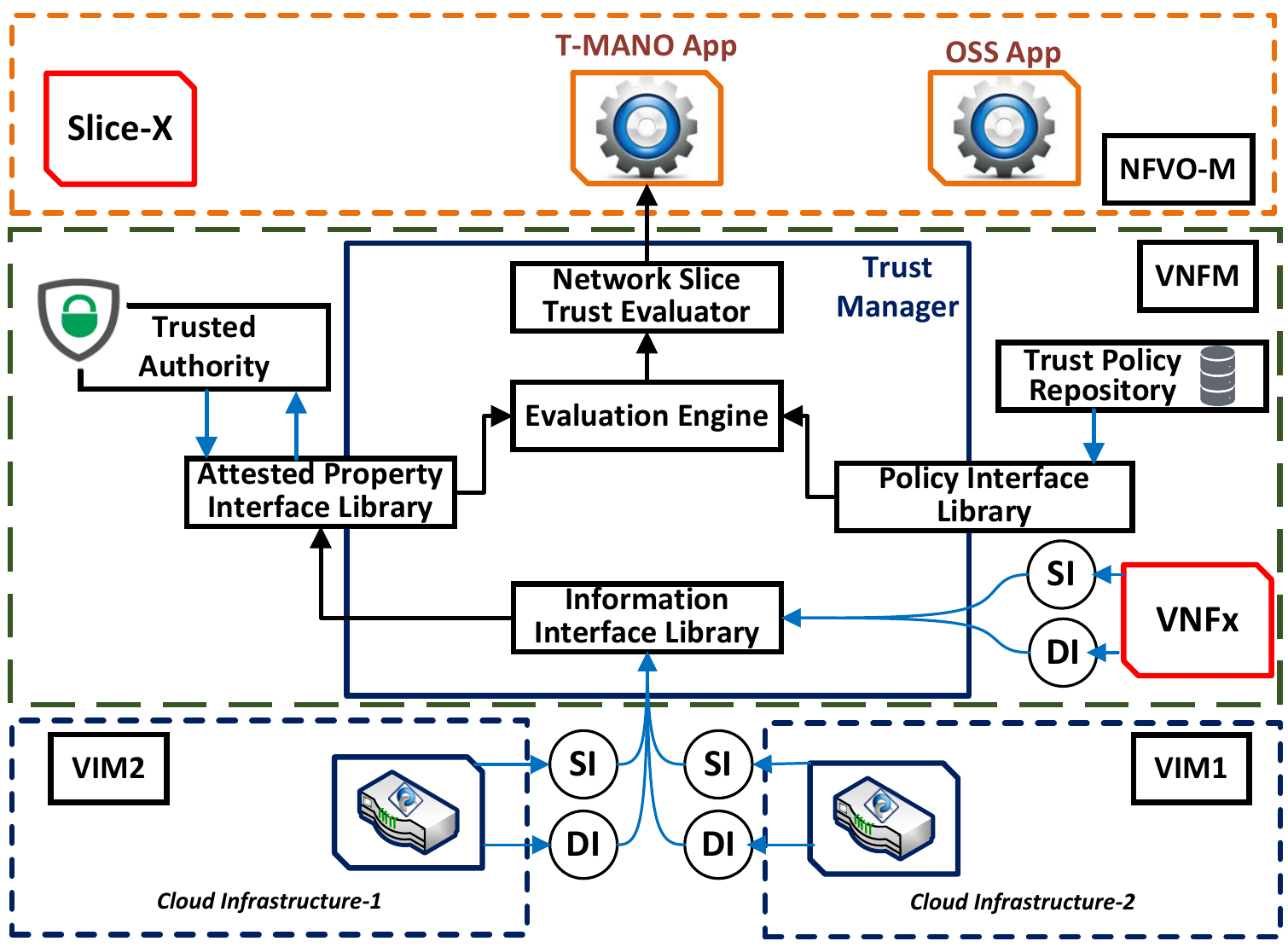}
    \caption{Trust Management Architecture}
    \label{fig:tmano}
\end{figure}
\noindent The four main components of the trust management architecture are: Trust Manager, T-MANO App, Trusted Authority and Trust Policy Repository. The Trust Manager is the primary component of the architecture. The T-MANO App is the front end interactive interface for the Trust Manager. The T-MANO App collects the users/providers' requests for trusted network slices. The Trust Manager evaluates the trust of the network slice which is composed of a set of VNFs deployed in multiple VMs across different physical platforms. The Trusted Authority determines the attested properties of the VNFs and VMs. In our architecture, it can perform both binary and property-based attestation. The Trust Policy Repository (TPR) specifies the trust requirements on the VNFS and VMs that need to be satisfied by the users and providers.\\
The basic operation of the trust management is as follows. The Trust Manager collects static and dynamic information about each VNF and the associated VMs where it is deployed. Static information is collected when the virtual machines are imported from the cloud infrastructure. Dynamic information consists of VNF deployment and VM run-time information. These happen via the Information Interface Library (IIL). It provides the static and dynamic information to the TA and obtains from the TA the corresponding properties (of the VNFs and VMs) attested by the TA. This happens via the Attested Property Interface Library (APIL). Then it retrieves the corresponding trust policies (for the VNFS and VMs) from the Trust Policy Repository (TPR) via the Policy Interface Library (PIL). The Evaluation Engine (EE) evaluates the trust of the VNFs and its associated VMs by checking whether the attested trust properties (from the TA) match with the required trust policies of VNFs and VMs (from the TPR). Finally, the overall trust of network slice is determined by aggregating the trust of the collection of VNFs and their associated VMs by the Network Slice Trust Evaluator (NSTE). This is then passed to the T-Mano App.
\vspace{-0.1cm}
\begin{figure}[!ht]
    \centering
   \includegraphics[scale=.5]{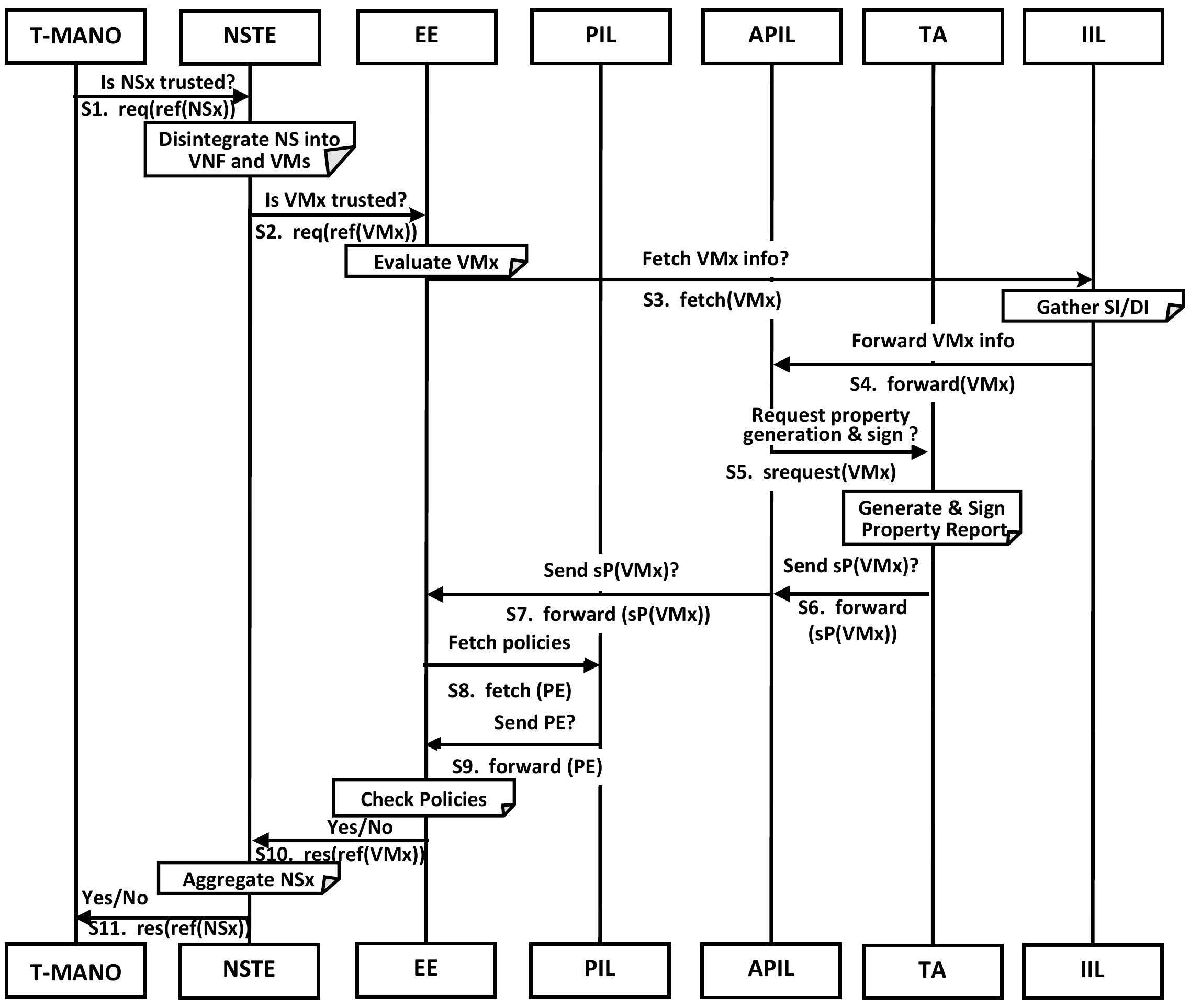}
   \caption{Timeline for Trust Evaluation of Network Slice}
    \label{fig:nstl}
\end{figure}

\noindent Let us now consider the operational view of the trust management architecture. \\
\noindent The architecture uses two types of information in its trust decision making.
\begin{definition}
\textbf{Static Information} - Information about virtual machines (VM) and VNFs when they are not actively participating in any network-related activity. \\
There are three occasions in which such information are generated: (i) when importing a VM file from the cloud infrastructure or a pre-configured VNF from a third-party storage; (ii) when deploying the VNF and associated VM and (iii) past information from a network slice that had been previously used by the MANO infrastructure but is currently sitting idle.
\end{definition} 
\begin{definition}
\textbf{Dynamic Information} - Information about a network slice and associated VNFs when they are actively participating in a network-related activity. For instance, a VNF being executed in VMs performing certain network activity.  
\end{definition} 
\noindent The overall operation is divided into two phases, namely, pre-deployment phase and deployment/active phase. The Trusted Authority (TA) acts as the attestation authority for both phases. It is capable of performing both binary and property-based attestation.  The sequence of events performed in both phases is the same. However, the types of information concerning network slices used are different. The pre-deployment phase-only uses static information, whereas the active phase uses both. Hence, we will describe only the active phase of the operation. \\
In our architecture, we have made the design choice to use a single TA for both VM and VNFM layers. In practice, this can be implemented as two separate authorities, one for the VNF and the other for the VM layer.\\
\noindent In the pre-deployment phase, VMs are imported from a third-party vendor in the cloud or can be created locally. In both cases, VIM generates VM hashes and checks them with the source hash. The IIL collects the physical platform and VM information. The APIL communicates with the TA for property-attestation purposes. The PIL collects the trust policies from the TPR. The EE evaluates the attested proerties against the policies. The NSTE is an aggregator module which combines the trust evaluation of the set of deployed VNFs on the VMs to determine the trustworthiness of the network slice. \\
Let us now consider the timeline of the events involved in the operation of the network slice trust evaluation as shown in Figure~\ref{fig:nstl}. A user or provider creates a network slice. The OSS App and VNFM help to create it. The VIM deploys the functions and VMs. There are two ways in which T-MANO App functions. First, it works in coordination with the OSS App, before and after the network slice deployment, to periodically check the trustworthiness of the network slices. Second, a user/provider at run-time can request the trustworthiness of a network slice.\\
Hence, we now consider how the T-MANO app requests the trust evaluation of the network slice, which is applicable to both these cases. Assume the T-MANO App asks the NSTE to evaluate the trustworthiness of a network slice ($S1$). It provides the Slice-ID as the identifier. The NSTE uses the VNFM store and network descriptor file to disintegrate the network slice information, to find out the individual VNFs and the VMs involved. It then asks the EE to evaluate each VNF/VM for their individual trustworthiness ($S2$).  The EE asks IIL to provide the static and dynamic information associated with the VNF/VM $S3$. The IIL maintains links with the cloud infrastructure hypervisor and VNFM to collect information about any specific VNF/VM (links are marked blue in the Figure). After collecting the information bundle, IIL forwards it to the APIL($S4$). The APIL uses this information bundle and asks TA for attestation($S5$). The TA checks the information and generates property attestation report (Listing 1.1 gives an example of a sample VNF Property Attestation Certificate).It signs this report and forwards it to the APIL($S6$). APIL forwards this to the EE($S7$).  The EE requests the PIL for the corresponding trust policies ($S8$). MANO administrators can add, update and delete trust policies in the TPR.(using a template based language). These policies are structured per domain and parameterized over VNF/VM and slice (Listing~\ref{prrepo} shows an example schema for this repository). The PIL forwards the specific policies of that network slice realm ($S9$). The EE evaluates the trust of each VNF/VM against these policies. It compares the attested trust properties of a VM or VNF against the trust policies in the TPR. It uses the LOPAT language and Algorithm 1 to check the properties. The VNF/VMs which satisfy the policies are considered as trusted. If they fail to satisfy the policies, they are considered as untrusted. In some cases, there can be conflicts as to refer any VNF/VM as trusted or untrusted; in such cases, they are termed as uncertain. The complete evaluation report is forwarded to the NSTE ($S10$).The NSTE then determines the overall trustworthiness of the specific Slice-ID. Finally, it provides the response to the query from the T-MANO App concerning the trustworthiness of the network slice with Slice-ID. If the network slice is untrustworthy, then the OSS App can dynamically replace the untrustworthy VNFs in that network slice and request for re-evaluation.    \\  
\vspace{-0.3cm}
\vspace{-0.3cm}
\section{Prototype Implementation and Analysis}
\vspace{-0.3cm}
\subsection{NFV Infrastructure Environment Setup}
\vspace{-0.2cm}
We have developed a proof of concept prototype of our trust management architecture. Our prototype uses one server (2 Intel Xeon CPU X5650 @2.67 GHz; 64GB of RAM) and a workstation (Intel Core i7 - 7700K @ 4.20GHz CPU; 64 GB of RAM) for the implementation. We have used Open Source MANO~\cite{etsi2016open} for creating the management infrastructure and OpenStack for the cloud infrastructure. We have deployed the Open Source MANO in the workstation. Open Source MANO is an ETSI standard open-source implementation of NFV Management and orchestration. OpenStack over the Xeon Server allows us to create and deploy IaaS cloud infrastructure enabling the creation, deployment and management of the VMs~\cite{pepple2011deploying}. We have used Python to develop different modules of our architecture. We have used JSON database for storing information about the properties. 
\vspace{-0.2cm}
\subsection{Performance Analysis}
\vspace{-0.3cm}
\begin{figure}[!ht]
    \centering
    \includegraphics[scale=.45]{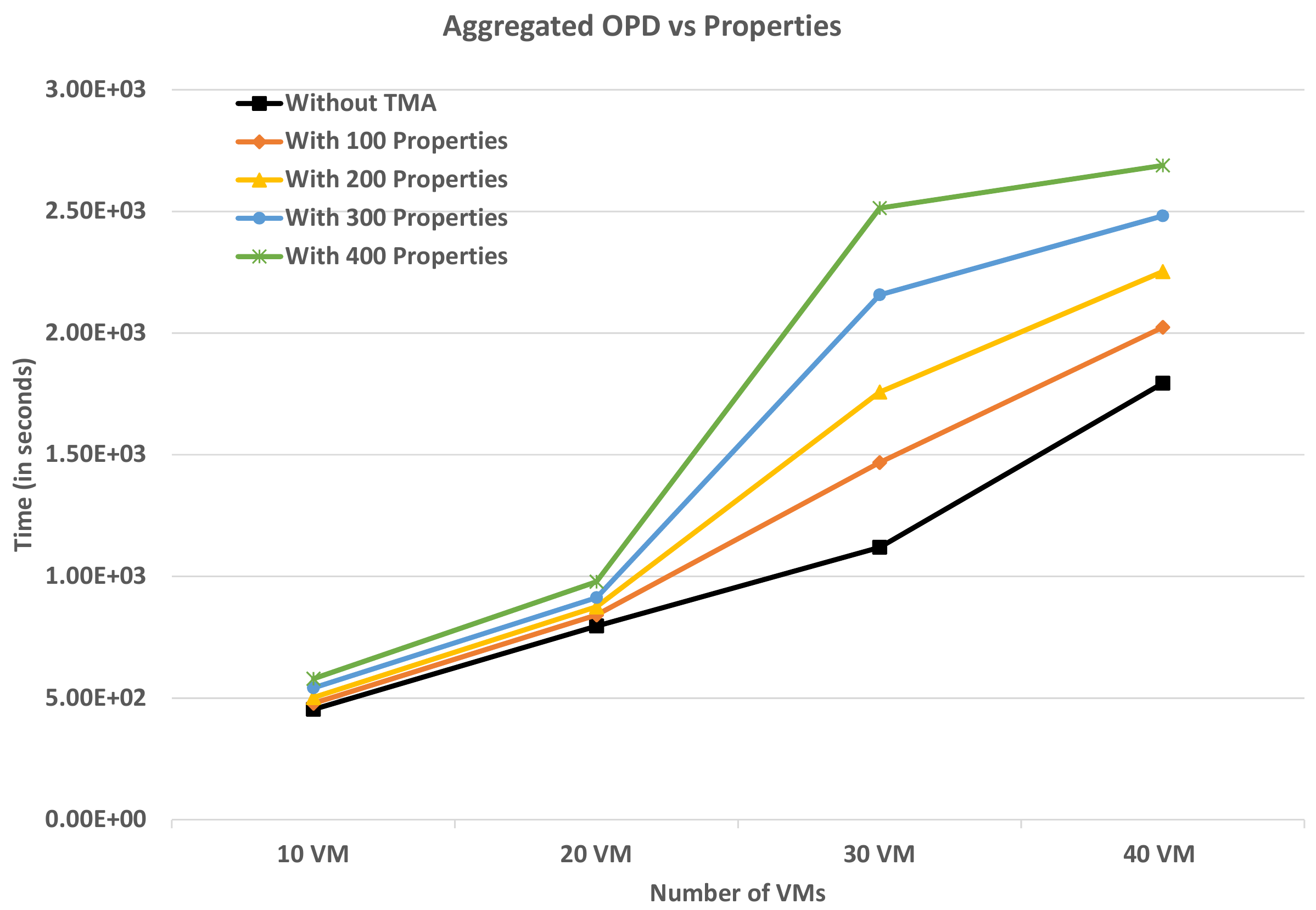}
    \caption{Aggregated On-boarding Processing Delay}
    \label{fig:opd}
\end{figure}
\vspace{.1cm}
\noindent On-boarding Process Delay (OPD) is the time required to boot a VNF and its associated VM resources~\cite{yilmabenchmarking}. The sum of such OPDs for an infrastructure is known as aggregated OPD. We have presented a comparison of the aggregated OPDs when using our trust management architecture in  Figure~\ref{fig:opd}. We have performed experiments to extract the delays. In these experiments, we have used a varying number of VMs from 10 to 40 to deploy the VNFs and measured the aggregated OPD time. In each case, we have varied the number of properties from 100 to 400 and measured the delays incurred. In order for the VMs to be trusted, they need to satisfy these properties. We have performed each experiment 10 times and have taken the average delay time. We have used Gnocchi to collect the OPDs~\cite{gno}. Figure~\ref{fig:opd} shows the variation of aggregated OPD with varying number of properties. One can see from the results that without our trust management  architecture, the aggregated OPD is $453.64$ seconds with 10 VMs and it increases to $1793.84$ seconds with 40 VMs. We will refer to this delay as the base delay for this specific network environment setup. With our trust management architecture, each VNF and associated VMs must go through the property-based attestation (As explained in the operation above) incurring a greater delay. For instance, with 10 VMs and 100 properties, the aggregated OPD measured from our experiments is $478.56$ seconds (which is approximately a $5$ percent increase). Over all our experiments, using our architecture, the aggregated OPD is higher in the range of $5-12$ percent compared to the base delay.\\
\begin{figure}[!ht]
    \centering
    \includegraphics[scale=.45]{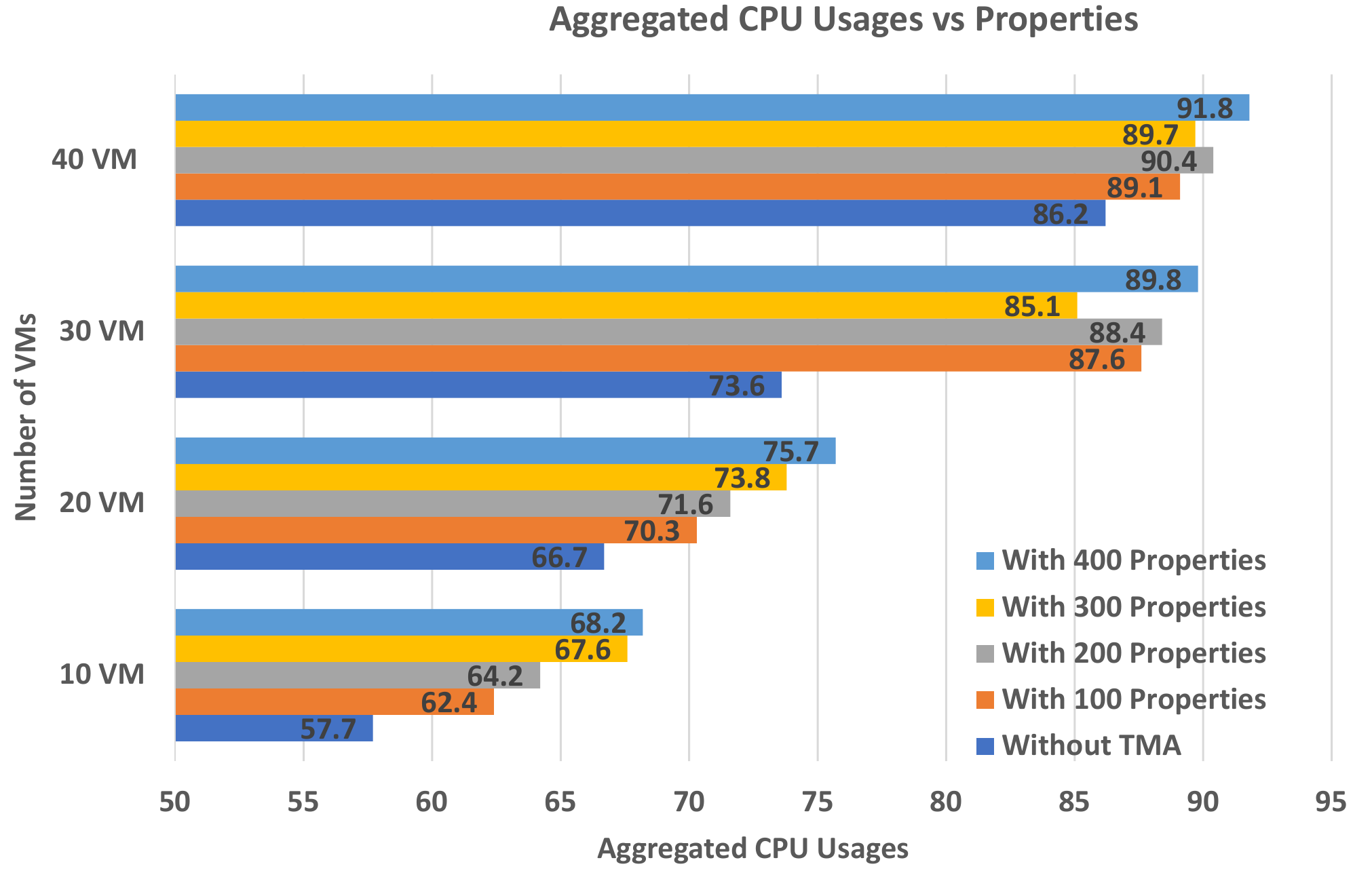}
    \caption{Aggregated CPU usages}
    \label{fig:cpu}
\end{figure}
The network slices and associated VNF/VMs consume a CPU resources and hence CPU usage is another important parameter to consider in performance analysis. Gnocchi helped us to measure the CPU usages. As the number of VMs increases, as expected, the CPU usage increases. For instance, the CPU usage was around $57.7$ percent with 10 VMs running and it increased to $86.2$ percent with 40 VMs. Using our trust management architecture, the CPU usage increases approximately 4-8 percent.   With 10 VMs using 400 properties, the aggregated CPU usage is $68.2$ percent and it increases to $91.8$ percent with 40 VMs. \\
We have carried out many experiments using different setups and have presented some of these results just to some indications of the increase in delay incurred due to the proposed architecture. The OpenMANO infrastructure performance is highly dependant on physical and virtual resources, and the configuration context of the network environment. Due to this diversity, there is not a standard performance benchmark for such systems~\cite{yilmabenchmarking}. Hence our performance metrics should not be treated as providing a standard but rather as some indications of the impact of our architecture on the overall performance. Furthermore, note that we have focused only on the trust related aspects of the performance. 
\vspace{-0.6cm}
\subsection{Attack Detection and Mitigation using Trust Architecture}
\vspace{-0.1cm}
In this section, we discuss an attack scenario specific to Mano infrastructure and show how our trust management architecture can help to detect and mitigate the attack.   \\
Consider a network service provider using a network slice X, which has VNF001 for processing data packets. The VNF001 is using two VMs to perform this network service. Assume both these VMs are Linux VMs and are manufactured by a third-party. \\
In the legitimate case, the network provider's slices will serve its users by processing their network packets, and the VMs will be doing the processing. \\
Assume that one of the Linux VMs ($192.168.56.109$) is equipped with \textit{zsh} shell. The VM also has the usual \textit{bash} shell. The \textit{zsh} shell is a legitimate shell and assume that it was not active during the time of deployment. Now consider this  specific VM has a logic bomb ($logicBOMB.sh$) script, which is time-triggered, meaning that it will become active after a certain amount of time elapses. Since the logic bomb script was inactive and \textit{zsh} shell was a legitimate shell, the integrity check during the VM deployment was unable to detect this anomaly. Also the property based trust attestation for the VM at that time turned out to be trusted.\\
Then we made the script alive, which activated the \textit{zsh} shell and disabled the address randomisation. This made the system vulnerable to buffer-overflow attack. Now, we introduced an adversary ($192.168.56.107$) to exploit the buffer-overflow to get a backdoor to the \textit{zsh} shell. Without our trust management solution, the adversary was successful in getting the backdoor to the VM and was able to launch further attacks. Figure~\ref{fig:sattack} shows such a successful attack. The upper part of the figure shows the $logicBOMB.sh$ script is running in the victim VM and the bottom part shows the reverse shell in the attackers machine. \\
However, our trust management solution is able to identify the changes in the VM and stop the attack. Recall our trust management architecture has a policy repository which has rules such as "there should not be any malicious script running in the VM". This policy must be satisfied by the VNF/VM of the network slice for it to execute. Once the script starts to execute, based on the dynamic information, the TA will not be able to provide a property based attestation satisfying the above rule in the trust policy. This will cause an alert to the T-MANO App, which in turn with the help of OSS App and VIM isolates the vulnerable VM. The vulnerable VM is then replaced with a suitable legitimate VM on the fly, thereby maintaining the activity of the network slice.
\vspace{-0.2cm}
\section{Related Work}
\vspace{-0.3cm}
The OpenMANO infrastructure uses a diverse number of platforms, hardware and software for the provision of network services as well as ensuring the quality of service demands of user and providers. Hence the need for trust relationships between the various components in the infrastructure and the services is critical. As far as we are aware the architecture proposed in this paper is one of the first ones to have gone through in detail the trust mechanisms required for the provision of trustworthy network slices. In this section, we will consider some related works that are relevant to various parts of our trust architecture, and discuss how our approach compares and extends these previous works.  \\
Paladi et al. in~\cite{paladi2018trust} presented a technique to secure the network assets like keys, certificates and session data in an isolated environment. They have used Intel -SGX~\cite{costan2016intel}to create trustworthy enclaves in OpenFlow devices which allows trusted network provisioning. This work focuses on SDN domain and OpenFlow devices. Our work extends this research by associating property-based trust attestation for the virtual and physical network assets in NFV infrastructures. Furthermore, our trust architecture caters for a multi authority-based infrastructure like MANO, which forms the foundation for 5G network architecture. \\
Trusted Click is another work that is relevant which uses Intel SGX based approach for NFV. Here the authors have used Click modular router~\cite{kohler2000click} to create secure NFV application enclaves. Each enclave will process and maintain the security and privacy of the network packets. For secure transmission of network packets from one SGX enclave to another, they have created secure SGX channels. In our proposed architecture, the focus is on the trustworthiness of the network slices at run time by using the trust properties of VNFs as well as those of VMs where the VNFs are deployed in the construction and provisioning of network slices. This enables to detect malicious behavioural changes in the VNFs alerting the management authorities as well as their dynamic replacement leading to continued provision of secure network services.  \\
TruSDN is another Intel-SGX based approach to bootstrap trust into SDN infrastructure~\cite{paladi2016trusdn}. TruSDN creates SGX enclaves for network endpoints and uses remote attestation to ensure their integrity. It allows TruSDN to establish secure communication channels between the endpoints. Our work extends this approach by performing both binary attestation and property-based attestation, both at boot and run time to ensure the trustworthiness of the network slices serving the user/providers. \\
Benedictis et al. in~\cite{de2017establishment} discusses issues in incorporating trust in ETSI specified MANO infrastructure. They have also demonstrated Intel Open Cloud Integrity Technology (CIT), which can be used in MANO to address certain trust issues.  However, this work does not consider the trustworthiness of a network slice, which we have addressed in this paper.\\
Bin et al. in~\cite{han2017security} propose a Trusted Zone architecture for the 5G network. Here, Trust Zones are defined as a set of network functions serving a local cell which has different policies to ensure data security. We believe the scope of trust policies is related to the notion of trust zone in their paper. Hence their work is somewhat complementary to our work. We can augment our work using their trust zone notion to define the domains associated with our trust policies. The trust architecture in this paper based on property based attestation, both at boot and run time, and the evaluation of trust of VNFs and VMs to determine the overall trustworthiness of a network slice all remain the same for a given domain.  \\
We have used property based attestation to develop trust enhanced authorisation systems for distributed services in our earlier works ~\cite{aarthi1}, ~\cite{aarthi2}. These have been relevant for the design of trust management architecture in this paper.
\vspace{-0.3cm}
\section{Concluding Remarks}
\vspace{-0.4cm}
Network slices in future communication networks (such as 5G) enable the provision of end to end logical networks running over a common infrastructure. They are needed for supporting a variety of vertical industries having a wide range of different service requirements. Security and trust issues play a critical role when there are different network slices operating over the same 5G infrastructure. \\
In this paper, we have addressed the fundamental issue of trust of a network slice. We have presented the design of a trust model to evaluate the trust of the of the virtual network functions that compose the network slice. The model evaluates the trust of the virtual network functions as well as the properties that should be satisfied by the virtual platforms (both at boot and run time) on which these network functions are deployed have for them to be trusted. It allows the users and providers to measure the trustworthiness of network slices.  \\
We have presented a logic-based language that defines simple rules for the specification of properties and the conditions under which these properties are evaluated to be satisfied for trusted virtualised functions and components.  We have designed a trust management architecture and mechanisms that enable the service providers to determine the trustworthiness of the network slices providing the network services.  We have described a prototype implementation of the trust management architecture using Open Source MANO Platform and presented the performance results. The results show that our trust mechanisms cause only a slight delay in the performance of the deployment of network slices establishing end to end services over virtualised infrastructures. We have also discussed how the proposed architecture can be used to detect and mitigate the impact of malicious virtual network functions in a dynamic manner.\\
Currently we are working on an enhanced version of the proposed architecture that can take into account reputation and recommendations from different network services and slices. We are also working on additional mechanisms that can be used to enhance the scalability of the proposed architecture.
\vspace{-0.3cm}
\bibliographystyle{splncs04}
\bibliography{ref}
\appendix
\section{Appendix: Attack Detection }
\begin{figure}[!ht]
    \centering
    \includegraphics[scale=.8]{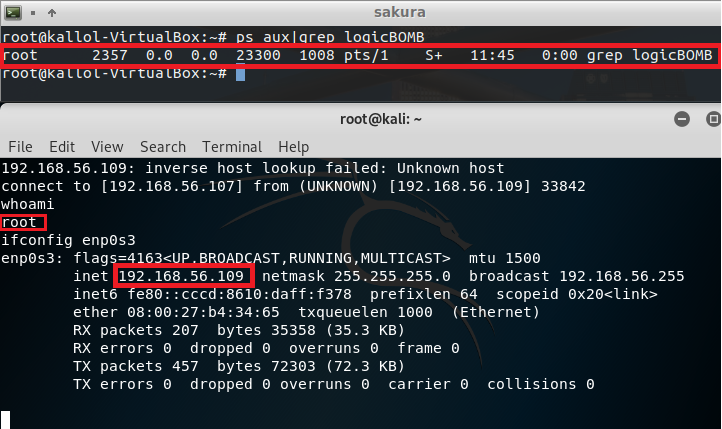}
    \caption{Successful Reverse Shell attack}
    \label{fig:sattack}
\end{figure}
\section{Appendix: Algorithms}
\begin{algorithm}[H]
 For each literal in Do\\
  \uIf{SatNS literal}{
    \uIf{SatNS derived from  PCert}{
        \uIf{prerequisites of SatNS are satisfied}{
        literal SatNS satified \;}
        }
    \uElseIf{SatNS in the head of NSP rule}{
        \uIf{Body of NSP rule contains SatNS literal}{
        Go to line 3\;
        }
        \uIf{Body of NSP rule contains HasNS literal}{
            \uIf{HasNS is derivable from DR}{
            Mark literal HasNS satified\;}
        }
        \uElseIf{Body of NSP rule contains SatC literal}{
        call CP resolution algorithm \;}
    }    
  }
  \uIf{all literals in Do are marked satisfied}{
  output permissions in Do\;}
\caption{Trust Derivation Algorithm}
\end{algorithm}
\begin{algorithm}[H]
Read the SatC literal in the head of the CP rule\\
  \uIf{SatC is derivable from PCert}{
    \uIf{prerequisites of SatC are Satisfied}{
    Mark literal SatC satisfied\;}
  }
  \uElseIf{SatC is head of another CP rule}{
    \uIf{Body of CP rule contains SatC literal}{
    Goto line 1\;}
    \uIf{Body of CP rules contains HasC literal}{
        \uIf{HasC is derivable from PCert}{
        Mark literal HasC satisfied\;}}
  }
\caption{CP Resolution Algorithm}
\end{algorithm}
\begin{algorithm}[H]\label{al:ev}
\LinesNumbered
\SetAlgoLined
 fetch\;
 \KwData{$VNF_x= {VNF_1, VNF_2, ..., VNF_x }$}
 \KwData{$SI_x= {SI_1, SI_2, ..., SI_N }$}
 \KwData{$DI_x= {DI_1, DI_2, ..., DI_N }$}
 \KwData{$Pr_i= {Pr_1, Pr_2, ..., Pr_n }$}
 \SetKwFunction{Fm}{evaluator}
 \SetKwFunction{s}{send}
 \SetKwFunction{Fps}{propertyStatus}
 \Fm{}\\
 {
 \ForEach{$VNF$ in $VNF_x$}{
     $status=$ \Fps{$SI_x$, $DI_x$, $Pr_i$} \\
    \SetAlgoLined
    \uIf{$status==untrusted$}
        {\s{isolate-VNF}\;
        $break$\;}
    \uElseIf{$status==trusted$}
        {\s{Trusted}\;
        $break$\;}
    \Else{\s{isolate-VNF}\;
        $break$\;}
 }
 }
 \Fps{$SI_x$,$DI_x$,$Pr_i$}{\\
    $T_D=convert(DI_x)$\;
    $T_S=convert(SI_x)$\;
    \ForEach{$Pr$ in $Pr_i$}{
    \SetAlgoLined
    \uIf{$Pr_i != (T_D\;AND\;T_S)$}
        {\Return $untrusted$\;}
    \uElseIf{$Pr_i == (T_D\;AND\;T_S)$}
        {\Return $trusted$\;}
    \Else{\Return $suspicious$\;
    }
    }
    }
%\EndFunction
 \caption{Evaluation Engine's Functional Algorithm }
\end{algorithm}
\section{Appendix: Repositories}
{\fontsize{6}{7}\selectfont
\begin{lstlisting}[caption=A Sample VNF Property Attestation Certificate,label=vmrepo]
<?xml version="1.0" encoding="UTF-8" ?>
<vnfCertificate>
	<certificateInfo>
		<id>00001</id>
		<issuer>TA</issuer>
		<issuerKey>"AAAAE2VjZHNhLXNoYTItbmlzdHAyNTYAAAAIbmlzdHAyNTYAAABBBE
		GfjPzZ1KEdoDAehYzGa+Upj0sYDV+Ol3OMg6em9vVxXjdCLcPmNq/vBYlTGqX/2g2uCNi7rD1DUPFI5CL1T3o"</issuerKey>
		<signAlgo>ECDSA</signAlgo>
		<digitalSign>a31193129ab9d43ed3f3014412773afb22ec6b1630e0931be0dac4fb653bd993</digitalSign>
		<validity>24hr</validity>
	</certificateInfo>
	<vnfInfo>
		<id>022RV</id>
		<vnfName>router</vnfName>
		<vnfMake>OF</vnfMake>
		<vnfPurpose>l2router</vnfPurpose>
		<vnfMap>
			<serviceVMinfo>
				<vmid>D1X022RV</vmid>
				<vmName>sakura</vmName>
				<vimLocation>"link"</vimLocation>
			</serviceVMinfo>
		</vnfMap>
	</vnfInfo>
	<staticProperty>
		<vnfHashinfo>
			<value>1a0f21437fc619acc51a81d552e9af77562263f7589f72752ac492caac9f7ed5</value>
			<issuer>ON</issuer>
			<type>SHA2</type>
		</vnfHashinfo>
		<serviceVMHashinfo>
			<value>d41dc6385e804fd6c6fe049ecd56a3c1bafa61e669d4f3b49082ff56f8ade10d</value>
			<issuer>ubuntu</issuer>
			<type>SHA2</type>
		</serviceVMHashinfo>
	</staticProperty>
	<dynamicProperty>
		<vnfProperty>
			<p>"No Malware"</p>
			<p>"Memory Integrity ok"</p>
			<p>"No Extra Service Running "</p>
		</vnfProperty>
		<serviceVMProperty>
			<p>"Trusted Processes are Running "
				<v>10</v>
			</p>
			<p>"No Memory Leakage"</p>
			<p>"No External Software Call"</p>
		</serviceVMProperty>
	</dynamicProperty>
</vnfCertificate>  
\end{lstlisting}
}
{\fontsize{6}{7}\selectfont
\begin{lstlisting}[caption=A Sample Trust Policy,label=prrepo]
<?xml version="1.0" encoding="UTF-8" ?>
<trustPolicy>
    <info>
        <id>01</id>
        <creator>Bob</creator>
        <cRole>admin</cRole>
    </info>
    <rule>
        <target>
            <platform> Any Network Slice</platform>
            <resources> Domain 1</resources>
        <condition>
            <entity>
                <vnf>
                    <sP> Hash is Valid </sP>
                    <sP> Digital Signature is Valid </sP>
                    <dP> No Malware </dP>
                    <dP> Memory Integrity OK  </dP>
                    <dP> No Extra Service Running </dP>
                </vnf> 
                <serviceVM>
                    <sP> Hash is Valid </sP>
                    <dP> No Memory Leakage </dP>
                    <dP> Trusted Processes are Running </dP>
                    <dP> No External Software Call </dP>
                </serviceVM>    
            </entity>
        </condition>
        <action> 
            <bTime>Trusted</bTime>
            <rTime>Trusted</rTime>
        </action>    
        </target>    
    </rule>
</trustPolicy>  
\end{lstlisting}
}

\end{document}